\documentclass[preprint,showpacs,aps,pra]{revtex4-1}
\usepackage{epsfig}
\usepackage{psfrag}
\usepackage{graphicx}
\usepackage{epstopdf}
\usepackage{graphics}
\usepackage{color}

\begin{document}


\title{Instabilities and propagation of neutrino magnetohydrodynamic waves in arbitrary direction}

\author{Fernando Haas and Kellen Alves Pascoal}
\affiliation{Instituto de F\'{\i}sica, Universidade Federal do Rio Grande do Sul, Av. Bento Gon\c{c}alves 9500, 91501-970 Porto Alegre, RS, Brasil}


\begin{abstract}
In a previous work \cite{NMHD}, a new model was introduced, taking into account the role of the Fermi weak force due to neutrinos coupled to  magnetohydrodynamic plasmas. The resulting neutrino-magnetohydrodynamics was investigated in a particular geometry associated with the magnetosonic wave, where the ambient magnetic field and the wavevector are perpendicular. The corresponding fast, short wavelength neutrino beam instability was then obtained in the context of supernova parameters. The present communication generalizes these results, allowing for arbitrary direction of wave propagation, including fast and slow magnetohydrodynamic waves and the intermediate cases of oblique angles. The numerical estimates of the neutrino-plasma instabilities are derived in extreme astrophysical environments where dense neutrino beams exist. 
\end{abstract}



\maketitle

\section{Introduction}

The neutrino-plasma coupling in magnetized media is a relevant issue in diverse situations, as near the core of proto-neutron stars, where it is a source of the free energy behind the stalled supernova shock \cite{Bludman}--\cite{Bethe2}. 
Neutrino-driven wakefields and neutrino effective charge in magnetized electron-positron plasma \cite{mag1, mag2}, the magnetized Mikheilev-Smirnov-Wolfenstein effect of neutrino flavor conversion \cite{mag3}, spin waves coupled to neutrino beams \cite{Semikoz},  neutrino cosmology and the early universe \cite{mag5}, neutrino emission and collective processes in magnetized plasma, and neutrino-driven nonlinear waves in magnetized plasmas \cite{mag8, mag9}, are examples of neutrino influenced plasma phenomena. The existence of intense neutrino beams in general astrophysical plasma is well documented \cite{Tajima}. The coupling between neutrino flavor oscillations and plasma waves has been also reported \cite{m1}--\cite{h2}. 

One of the most popular approaches to plasma astrophysics in the presence of magnetic fields is magnetohydrodynamics (MHD), which usually does not account for neutrino species not even in any approximate way. Actually, neutrino studies in a material medium are more frequently pursued within the framework of particle physics, which in terms of language is somewhat far from the majority of the plasma community. This has motivated the creation of neutrino-magnetohydrodynamics (NMHD), where the interaction between neutrinos and electrons is forwarded in terms of a coupling between the MHD and neutrino fluids \cite{NMHD}. As a first application, NMHD proved the destabilization of the magnetosonic wave by neutrino beams, yielding a plausible mechanism for type II supernova explosion. However, the magnetosonic wave supposes a very particular geometry, where the wave propagation is perpendicular to the ambient magnetic field. Therefore, it is advisable to perform a more general linear stability analysis, allowing for arbitrary orientations. This is the purpose of the present work, namely, the study of the impact of a neutrino beam on the stability of general MHD waves. Namely, in the case of an ideally conducting fluid and using simplified MHD assumptions, these are the shear Alfv\'en wave, and fast and slow magnetosonic waves. Therefore, the present work removes the orthogonality condition of \cite{NMHD}, to obtain instability growth-rates of simplified and ideal NMHD for arbitrary oblique angles between wave propagation and equilibrium magnetic field. Similarly, the instability analysis of general electrostatic perturbations in magnetized electron plus neutrino plasmas in an ionic background was recently carried on \cite{PRD}. 

It can be justifiable argued that the NMHD model as it stands underestimates other important quantum effects in dense plasmas, such as relativistic degeneracy effects, particle dispersive effects and exchange effects \cite{book}. The basic reason for our choice is that the original quantum magnetohydrodynamics was derived starting from a quantum kinetic model, the non-relativistic Wigner-Maxwell system, not including neutrino coupling \cite{qmhd}. Therefore the insertion of relativistic corrections and extra terms of exchange and quantum dispersion would be {\it ad hoc} in the present state of the art. On the other hand, for very dense white dwarfs, degeneracy comes together with relativistic effects in view of a Fermi momentum $p_F$ of the order of $mc$, where $m$ is the mass of the charge carriers and $c$ the speed of light. Hence for strongly degenerate-relativistic plasmas  a more advanced theory would be necessary from the beginning.

This work is organized as follows. Section II reviews the basic equations and validity conditions of NMHD. Section III obtains the general linear dispersion of waves, where a few extra details (not explicitly shown in \cite{NMHD}) of the algebra are provided. Section IV derives the instability growth-rate in general, discussing it in the significant particular cases: fast magnetosonic wave; slow magnetosonic wave; perpendicular wave propagation (with respect to the ambient magnetic field); parallel wave propagation. The shear Alfv\'en wave is found to be unaffected by neutrinos. The strong growth-rate is estimated in a typical case of type II supernova parameters. Section V is reserved to the conclusions. 

\section{Neutrino-magnetohydrodynamics model}

For completeness, we briefly review the NMHD model derived in \cite{NMHD}, comprising the following set of equations, namely, the continuity equations for the neutrinos, 
\begin{equation}
 \frac{\partial n_\nu}{\partial t} + \nabla \cdotp (n_\nu \textbf{u}_\nu) = 0 \,, \label{eq32} 
\end{equation}
and for the MHD fluid, 
\begin{equation}
 \frac{\partial \rho_m}{\partial t} + \nabla \cdot (\rho_m \textbf{U}) = 0 \,,  \label{eq26} 
\end{equation}
the momentum transport equations for the neutrinos,
\begin{equation}
 \frac{\partial \textbf{p}_\nu}{\partial t} + \textbf{u}_\nu \cdotp \nabla \textbf{p}_\nu = - \frac{\sqrt{2}\,G_F}{m_i} \nabla \rho_m \,, \label{nf} 
\end{equation}
and for the MHD fluid, 
\begin{equation}
 \frac{\partial \textbf{U}}{\partial t} + \textbf{U} \cdot \nabla \textbf{U} = - \frac{V_{S}^2 \nabla \rho_m}{\rho_m} + \frac{(\nabla\times\textbf{B}) \times \textbf{B}}{\mu_0 \,\rho_m} + \frac{\textbf{F}_\nu}{m_i} \,, \label{eq34} 
\end{equation}
as well as the dynamo equation modified by the electroweak force, 
\begin{equation}
 \frac{\partial\textbf{B}}{\partial t} = \nabla\times\left(\textbf{U}\times\textbf{B} - \frac{\textbf{F}_{\nu}}{e}\right) \,. \label{eq37}
\end{equation}
Here, $n_\nu$ and $\rho_m$ are resp. the neutrino number density and the plasma mass density, ${\bf u}_\nu$ and ${\bf U}$ resp. the neutrino and plasma velocity fields, ${\bf B}$ the magnetic field, $G_F$ the Fermi constant, $m_i$ the ion mass, $V_S$ the adiabatic speed of sound, $\mu_0$ the free space permeability, $e$ the elementary charge and ${\bf F}_\nu$ the neutrino force,   
\begin{equation}
 \textbf{F}_\nu = \sqrt{2}\,G_F \left[\textbf{E}_\nu + \left(\textbf{U} - \frac{m_i \nabla\times{\bf B}}{e \mu_0 \rho_m}\right) \times \textbf{B}_\nu \right] \,, \label{eq29}
\end{equation}
where $\textbf{E}_{\nu}$ and $\textbf{B}_{\nu}$ are effective fields induced by the weak interaction, 
\begin{eqnarray}
 \textbf{E}_\nu = \nabla n_{\nu} - \frac{1}{c^2}\frac{\partial}{\partial t}(n_{\nu} \textbf{u}_{\nu}) \,, \quad \textbf{B}_\nu = \frac{1}{c^2}\nabla  \times (n_{\nu}\textbf{u}_{\nu}) \,. \label{eq04}
\end{eqnarray}
Finally, the neutrino relativistic beam momentum is ${\bf p}_\nu =  \mathcal{E}_{\nu} \textbf{u}_{\nu}/c^2$, with a neutrino beam energy $\mathcal{E}_{\nu}$. 

The assumptions behind the NMHD model are the same of the simplified and ideal MHD, namely, a highly conducting, strongly magnetized medium, and low frequency processes in a scale where electrons and ions couple so much as to be faithfully treated as a single fluid. The neutrinos influence the plasma by means of the charged weak current coupling electrons and electron-neutrinos, through the
charged bosons $W_{\pm}$. In addition, implicitly in Eq. (\ref{eq34}) the displacement current was neglected, supposing wave phase velocities much smaller than $c$ - although such a restriction has no r\^ole in the results of the present work. In conclusion, Eqs. (\ref{eq32})-(\ref{eq37}) are a complete set of 11 equations and 11 variables, namely $n_\nu, \rho_m$ and the components of $\textbf{p}_\nu, \textbf{U}$ and $\textbf{B}$. A more detailed derivation is provided in \cite{NMHD}. 

For convenience, it is useful to reproduce here Eq. (28) of \cite{NMHD}, which collects the conditions of high collisionality and high conductivity of the plasma, supposing a wave with angular frequency $\omega$, 
\begin{equation}
\frac{m_i |\omega|}{m_e \omega_{pe}} \ll \frac{2}{3}\,\frac{\ln\Lambda}{\Lambda} \ll \frac{\omega_{pe}}{|\omega|} \,, \quad \Lambda = \frac{4\pi n_0 \lambda_D^3}{3} \,, \quad \lambda_D = \frac{v_T}{\omega_{pe}} \,, \label{con}
\end{equation}
where $n_0$ is the equilibrium electron (and ion) number density, $m_e$ is the electron mass, $\omega_{pe} = [n_0 e^2/(m_e \varepsilon_0)]^{1/2}$ is the electron plasma frequency, $v_T = (\kappa_B T_e/m_e)^{1/2}$ is the electrons thermal velocity, $\kappa_B$ is the Boltzmann constant and $T_e$ the electron fluid temperature. The validity conditions of NMHD are essentially the same, since the neutrino component is a second order influence. The derivation of Eq. (\ref{con}) assumes the Landau electron-electron collision frequency, and non-degenerate and non-relativistic electrons. More details on the validity conditions of MHD can be found e.g. in \cite{Spitzer, Balescu}.

\section{General dispersion relation}
Starting from the homogeneous equilibrium 
\begin{eqnarray}
 \quad n_{\nu} = n_{\nu 0} \,, \quad \rho_m = \rho_{m0} \,,  \quad \textbf{p}_\nu = \textbf{p}_{\nu 0} \,, \quad \textbf{U} = 0 \,, \quad \textbf{B} = \textbf{B}_0 \,,
\end{eqnarray}
and supposing plane wave perturbations proportional to $\exp[i({\bf k}\cdot{\bf r} - \omega\,t)]$, it is possible to obtain the dispersion relation for small amplitude waves. Here we provide a few more details on the necessary algebra, in comparison with \cite{NMHD}. The idea is to express all perturbations in terms of $\delta{\bf U}$, the first-order plasma fluid correction. For instance, the linear correction to the neutrino fluid velocity becomes 
\begin{eqnarray}
\delta{\bf u}_\nu &=& \frac{c^2}{\mathcal{E}_{\nu 0}}\left(\delta{\bf p}_\nu - {\bf u}_{\nu 0}\,{\bf u}_{\nu 0}\cdot\delta{\bf p}_\nu/c^2\right) \label{eqax} \\
&=& \frac{\sqrt{2} G_F \rho_{m\,0} c^2}{m_i \mathcal{E}_{\nu 0}\,\omega}\,\frac{\left({\bf k} - {\bf k}\cdot{\bf u}_{\nu 0}\,{\bf u}_{\nu 0}/c^2\right)}{(\omega - {\bf k}\cdot{\bf u}_{\nu 0})}\,\,{\bf k}\cdot\delta{\bf U} \,, \label{eqxx}
\end{eqnarray}
where ${\bf u}_{\nu 0}$ and $\mathcal{E}_{\nu 0}$ are resp. the equilibrium neutrino beam velocity and energy, viz. ${\bf p}_{\nu 0} =  \mathcal{E}_{\nu 0} \textbf{u}_{\nu 0}/c^2$. Equation (\ref{eqax}) can be operationally found using the relation between neutrino momentum and neutrino velocity and the energy-momentum relation $\mathcal{E}_{\nu} = (p_\nu^2 c^2 + m_\nu^2 c^4)^{1/2}$, where the neutrino mass $m_\nu$ is eliminated at the end. The step from Eq. (\ref{eqax}) to Eq. (\ref{eqxx}) is made using the linearized plasma continuity equation (\ref{eq26}) and the linearized neutrino momentum transport equation (\ref{nf}). 

To proceed, in view of Eq. (\ref{eq29}) the linearized neutrino force becomes $\delta{\bf F}_\nu = \sqrt{2} G_F \delta{\bf E}_\nu$ since the term containing the effective neutrino magnetic field ${\bf B}_\nu$ is of second order. The perturbed effective neutrino electric field $\delta{\bf E}_\nu$ can be found from Eq. (\ref{eq04}), together with the neutrino continuity equation (\ref{eq32}) and Eq. (\ref{eqxx}). The result is 
\begin{eqnarray}
\delta{\bf F}_\nu &=& \frac{2 i G_F^2 n_{\nu 0}\rho_{m0}\,({\bf k}\cdot\delta{\bf U})}{m_i \mathcal{E}_{\nu 0}\,\omega (\omega - {\bf k}\cdot{\bf u}_{\nu 0})^2} \times \nonumber \\ &\times& \Bigl[\Bigl(({\bf k}\cdot{\bf u}_{\nu 0})^2 - c^2 k^2 - \omega ({\bf k}\cdot{\bf u}_{\nu 0}) + \omega^2\Bigr)\,{\bf k} + \omega\,\Bigl(k^2 - \frac{\omega}{c^2}\,{\bf k}\cdot{\bf u}_{\nu 0}\Bigr)\,{\bf u}_{\nu 0}\Bigr] \,.
\end{eqnarray}
As could have been expected, the neutrino force is enhanced for $\omega \approx {\bf k}\cdot{\bf u}_{\nu 0}$, so that the wave resonates with the neutrino beam. 

The remaining straightforward steps allow to express the linearized plasma momentum transport equation (\ref{eq34}) in terms of $\delta{\bf U}$ only, 
\begin{eqnarray}
 \omega^2\delta\textbf{U} &=& \left(V^2_A + V^2_S + V^2_N \, \Bigl(\frac{c^2k^2 - (\textbf{k}\cdot\textbf{u}_{\nu 0})^2 + \omega ({\bf k}\cdot{\bf u}_{\nu 0}) - \omega^2}{(\omega- \textbf{k}\cdot \textbf{u}_{\nu 0})^2}\Bigr)\right)\!(\textbf{k}\cdot\delta\textbf{U})\textbf{k} \nonumber \\ &+& (\textbf{k} \cdot \textbf{V}_A)\Bigl((\textbf{k} \cdot \textbf{V}_A)\delta\textbf{U} - (\delta\textbf{U}\cdot\textbf{V}_A)\textbf{k} \nonumber - (\textbf{k}\cdot\delta\textbf{U})\textbf{V}_A\Bigr)   \nonumber \\
&-& \frac{\omega\, V_N^2 \Bigl(k^2 - \omega {\bf k}\cdot{\bf u}_{\nu 0}/c^2\Bigr)({\bf k}\cdot\delta{\bf U})\,{\bf u}_{\nu 0}}{(\omega- \textbf{k}\cdot \textbf{u}_{\nu 0})^2} \nonumber \\ 
&+& \frac{i V_N^2 V_A (\textbf{k}\cdot\delta\textbf{U})}{\Omega_i (\omega- \textbf{k}\cdot \textbf{u}_{\nu 0})^2}\,\Bigl(k^2 - \frac{\omega\, {\bf k}\cdot{\bf u}_{\nu 0}}{c^2}\Bigr)\,{\bf V}_A \times \Bigl({\bf k}\times ({\bf k}\times{\bf u}_{\nu 0})\Bigr) \,, \label{ccc}
\end{eqnarray}
where the vector Alfv\'en velocity ${\bf V}_A$ and $V_N$ are given by 
\begin{eqnarray}
\textbf{V}_A = \frac{\textbf{B}_0}{(\rho_{m0} \mu_0)^{1/2}} \,, \quad 
V_N = \left(\frac{2G^2_F \rho_{m0} n_{\nu0}}{m^2_i \mathcal{E}_{\nu 0}}\right)^{1/2} \,, \label{eq53}
\end{eqnarray}
while $\Omega_i = e B_0/m_i$ is the ion cyclotron frequency. As apparent, the characteristic neutrino-plasma speed $V_N$ contains both MHD and neutrino variables, emphasizing the mutual coupling. 

The somewhat formidable expression can be considerably simplified for low frequency waves such that $\omega/k \ll c$, allowing to disregard the terms containing $\omega$ in the numerators of the right-hand side of Eq. (\ref{ccc}), as deduced from appropriated order of magnitude estimates. In the same trend, the very last term proportional to $\Omega_{i}^{-1}$ can be discarded, provided $k V_A/\Omega_i \ll c/V_A$, or equivalently $c k/\omega_{pe} \ll \omega_{pe}/\Omega_e$, where $\Omega_e = e B_0/m_e$ is the electron cyclotron frequency. Such a condition tend to be easily satisfied wavelengths much larger than the plasma skin depth $c/\omega_{pe}$, and large enough densities so that $\omega_{pe} \gg \Omega_e$. Finally, Eq. (\ref{ccc}) reduces to 
\begin{eqnarray}
 \omega^2\delta\textbf{U} &=& \left(V^2_A + V^2_S +   V^2_N \frac{(c^2k^2 - (\textbf{k}\cdot\textbf{u}_{\nu 0})^2)}{(\omega- \textbf{k}\cdot \textbf{u}_{\nu 0})^2}\right)\!(\textbf{k}\cdot\delta\textbf{U})\textbf{k} \nonumber \\ &+& (\textbf{k} \cdot \textbf{V}_A)\Bigl((\textbf{k} \cdot \textbf{V}_A)\delta\textbf{U} - (\delta\textbf{U}\cdot\textbf{V}_A)\textbf{k} - (\textbf{k}\cdot\delta\textbf{U})\textbf{V}_A\Bigr)  \,, \label{eq51}
\end{eqnarray}
which is shown in \cite{NMHD}. 

In \cite{NMHD}, for simplicity it was supposed that ${\bf k}\cdot{\bf V}_A = 0$, which allows to discard several terms of  Eq. (\ref{eq51}). This corresponds to the magnetosonic wave modified by the neutrino component, for which $\delta{\bf U} \parallel {\bf k}$ as seen from inspection. The corresponding instability due to the neutrino beam was then evaluated. Our goal now is to consider the general situation, where the wavevector and the ambient magnetic field have an arbitrary orientation, as shown in Fig. 1. 

\begin{figure}[h]
\begin{center}
\includegraphics[width=2.5in]{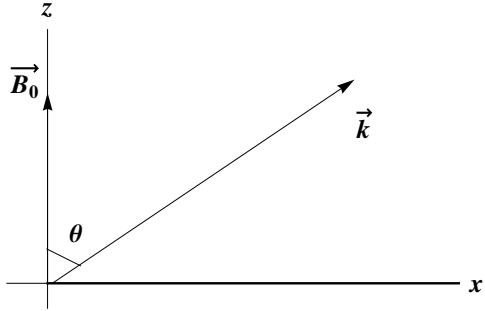}
\caption{Wave vector and ambient magnetic field.}
\label{fig1} 
\end{center}
\end{figure}

It turns out that Eq. (\ref{eq51}) is formally the same as the one for linear waves in simplified ideal MHD, provided the adiabatic sound speed $V_S$ is replaced by $\tilde V_S(\omega,{\bf k})$ defined by 
\begin{equation}
 \widetilde{V}_S^2(\omega,{\bf k}) = V_S^2 + V_N^2 \frac{(c^2k^2 - (\textbf{k}\cdot \textbf{u}_{\nu 0})^2)}{(\omega - \textbf{k} \cdot \textbf{u}_{\nu 0})^2} \,,
\end{equation}
so that 
\begin{eqnarray}
 \omega^2\delta\textbf{U} &=& \left(V^2_A + \tilde{V}^2_S(\omega,{\bf k})\right)\!(\textbf{k}\cdot\delta\textbf{U})\textbf{k} \nonumber \\ &+& (\textbf{k} \cdot \textbf{V}_A)\Bigl((\textbf{k} \cdot \textbf{V}_A)\delta\textbf{U} - (\delta\textbf{U}\cdot\textbf{V}_A)\textbf{k} - (\textbf{k}\cdot\delta\textbf{U})\textbf{V}_A\Bigr)  \,, \label{eqq51}
\end{eqnarray}
which is exactly the same as the well known simplified and ideal MHD system for linear waves, with the replacement $V_S \rightarrow \tilde{V}_S(\omega,{\bf k})$. Hence, the usual procedure applies, as follows. 

Assuming the geometry of Fig. 1, where without loss of generality the $y-$component of ${\bf k}$ and ${\bf V}_A$ is set to zero, and from the characteristic determinant of the homogeneous system (\ref{eqq51}) for the components of $\delta{\bf U}$, the result is 
\begin{equation}
(\omega^2 - k^2\,V_A^{~2}\cos^2\theta)\left[ \omega^4 -  k^2 \left(V_A^2+ \widetilde{V}_S^{~2}(\omega,{\bf k})\right)\omega^2 + k^4\,V_A^{~2}\,\widetilde{V}_S^{~2}(\omega,{\bf k})\,\cos^2\theta \right] = 0 \,. \label{sa}
\end{equation}

As apparent from the factorization, one root is $\omega = k\,V_A\,\cos\theta$, which is the shear Alfv\'en wave, unaffected by the neutrino beam. This happens because ${\bf k}\cdot\delta{\bf U} = 0$ for the shear Alfv\'en wave, which eliminates the neutrino contribution in Eq. (\ref{eqq51}). 
Presently, the more interesting modes comes from the second bracket in Eq. (\ref{sa}), to be discussed in the next Section.   

\section{Instabilities} 

Ignoring the shear Alfv\'en wave, the general dispersion relation (\ref{sa}) yields
\begin{equation}
\label{new}
\omega^4 -  k^2 (V_A^2+ V_S^{~2})\omega^2 + k^4\,V_A^{~2}\,V_S^{~2}\,\cos^2\theta = \frac{V_N^2 k^2 \left(c^2 k^2 - ({\bf k}\cdot{\bf u}_{\nu 0})^2\right) (\omega^2 - k^2 V_A^2 \cos^{2}\theta)}{(\omega - {\bf k}\cdot{\bf u}_{\nu 0})^2} \,,
\end{equation}
where the neutrino term was isolated in the right-hand side. Due to the small value of the Fermi constant, the neutrino contribution is always a perturbation, even for the neutrino-beam mode. The natural approach to Eq. (\ref{new}) is then to set 
\begin{equation}
\label{xx}
\omega = \Omega + \delta\omega \,, \quad \Omega \gg \delta\omega \,, \quad \Omega = {\bf k}\cdot{\bf u}_{\nu 0} \,,
\end{equation}
where $\Omega$ is the classical (no neutrinos) solution, 
\begin{equation}
\label{nwew}
\Omega^4 -  k^2 (V_A^2+ V_S^{~2})\Omega^2 + k^4\,V_A^{~2}\,V_S^{~2}\,\cos^2\theta = 0 \,,
\end{equation}
and where in Eq. (\ref{xx}) the neutrino-beam mode was selected in order to enhance the neutrino contribution.

Therefore, the zeroth-order solution gives the fast (+) and slow (-) magnetosonic waves, 
\begin{equation} \Omega = \Omega_{\pm} = k V_{\pm} \,, \quad V_\pm = \left[\frac{1}{2}\left(V_A^{~2} + V_S^{~2} \pm \sqrt{(V_A^2 - V_S^2)^2 + 4\, V_A^{~2}\,V_S^{~2}\,\sin^2\theta} \right)\right]^{1/2} \,. \label{vpm}
\end{equation}

Taking into account Eq. (\ref{new}) and Eq. (\ref{xx}) as well as the expression of the unperturbed frequency, we get
\begin{eqnarray}
(\delta\omega)^3 &=&  \frac{\pm V_N^2 \left(c^2 k^2 - ({\bf k}\cdot{\bf u}_{\nu 0})^2\right) \left(V_\pm^2 - V_A^2 \cos^{2}\theta\right) k}{2 V_\pm \, \sqrt{(V_A^2 - V_S^2)^2 + 4\, V_A^{~2}\,V_S^{~2}\,\sin^2\theta} } \nonumber \\ 
&\approx&
 \frac{\pm V_N^2 c^2 \left(V_\pm^2 - V_A^2 \cos^{2}\theta\right) k^3}{2 V_\pm \, \, \sqrt{(V_A^2 - V_S^2)^2 + 4\, V_A^{~2}\,V_S^{~2}\,\sin^2\theta}} \,, 
\end{eqnarray}
where in the last step $\Omega = {\bf k}\cdot{\bf u}_{\nu 0}$ and $V_{\pm}^2 \ll c^2$ were used. The unstable root with $\gamma = {\rm Im}(\delta\omega) > 0$ yields the growth-rate 
\begin{equation}
\gamma = \gamma_\pm = \frac{\sqrt{3}\,k}{2^{4/3}} \left(\frac{\Delta c^4 |V_{\pm}^2 - V_A^2 \cos^{2}\theta|}{V_{\pm} \, \sqrt{(V_A^2 - V_S^2)^2 + 4\, V_A^{~2}\,V_S^{~2}\,\sin^2\theta} }\right)^{1/3} \,, 
\label{gamma}
\end{equation}
introducing the dimensionless quantity
\begin{equation}
\Delta = \frac{V_N^2}{c^2} = \frac{2G_F^2n_0n_{\nu 0}}{m_i c^2E_{\nu 0}} \,,
\end{equation}
using $\rho_{m0} \approx n_0 m_i$. The parameter $\Delta$ is endemic in neutrino-plasma problems, as in the neutrino and anti-neutrino effective charges in magnetized plasmas \cite{mag1} or in the expression of the neutrino susceptibility \cite{Silva}.

The weak beam condition $\gamma/\Omega \ll 1$ can be worked out as
\begin{equation}
\frac{\Delta c^4 |V_{\pm}^2 - V_A^2 \cos^{2}\theta|}{V_{\pm}^4 \,\sqrt{(V_A^2 - V_S^2)^2 + 4\, V_A^{~2}\,V_S^{~2}\,\sin^2\theta} } \ll 1 
\,, \label{weak}
\end{equation}
which is independent of the magnitude $k$ of the wavenumber. In the unlikely cases where  Eq. (\ref{weak}) is not satisfied,  one must go back to the sixth-order polynomial equation (\ref{new}), to be numerically solved.   

The growth-rate (\ref{gamma}) is completely general, in the sense that it is valid for arbitrary geometries of the wave propagation, as long as the weak beam assumption holds, and is the main result of this work. It is interesting to evaluate the instability in the separate fast and slow magnetosonic cases, as well as for perpendicular (${\bf k} \perp {\bf V}_A$) and parallel (${\bf k} \parallel {\bf V}_A$) to the magnetic field wave propagation. 

\subsection{Destabilization of the fast magnetosonic wave}  
The choice of the plus sign in Eq. (\ref{gamma}) corresponds to the fast magnetosonic wave, with a growth-rate $\gamma \equiv \gamma_+$. From now on, parameters of Type II core-collapse scenarios like for the supernova SN1987A will be applied. There one had neutrino bursts of $10^{58}$ neutrinos and energies of the order of $10-15$ MeV, strong magnetic fields $B_0 \approx 10^6 - 10^8 \,{\rm T}$ and neutrino beam densities $n_{\nu 0}$ between $10^{34} - 10^{37} \, {\rm m}^{-3}$ \cite{Hirata}. In the following estimates, we set ${\cal E}_{\nu 0} = 10 \,{\rm MeV}, n_0 = 10^{34}\,{\rm m}^{-3}$, $n_{\nu 0} = 10^{35} \,{\rm m}^{-3}$, $B_0 = 5 \times 10^7 \,{\rm T}$, and an electron fluid temperature $T_e = 0.1 \, {\rm MeV}$, appropriate for the slightly degenerate and mildly relativistic hydrogen plasma in the center of the proto-neutron star. In addition, we use $G_F=1.45 \times 10^{-62} \, {\rm J.m^3}$, $V_S = (\kappa_B T_e/m_i)^{1/2}$. For these parameters, one has $\Delta = 1.75 \times 10^{-33}$, $V_A/c = 3.64 \times 10^{-2}, V_S/c = 1.03 \times 10^{-2}$. We set $k = 10^{6} \,{\rm m}^{-1}$, which is fully consistent with  the applicability condition (\ref{con}). Finally, the simplifying assumption of page 6, viz. $c k/\omega_{pe} \ll \omega_{pe}/\Omega_e$, becomes $k \ll 1.2 \times 10^{10}\,{\rm m}^{-1}$, which is obviously satisfied. 

From Eq. (\ref{gamma}), the result is then shown in Fig. 2, displaying the growth-rate as a function of the orientation angle. One has a fast instability with the estimate $1/\gamma_+ \approx 10^{-3} \,{\rm s}$, while the characteristic time of supernova explosions is $\sim$ 1 second. On the other hand, the weak beam assumption $\gamma_+ \ll \Omega_+$ (equivalent to Eq. (\ref{weak})) is fairly satisfied, since $\Omega_+ \approx 10^{13} \,{\rm rad/s}$ without much variation as a function of the angle. The conclusion from Fig. 2 is that the instability becomes stronger for more perpendicular waves. One could have even stronger instabilities for a denser plasma, but some of the above calculations, although remaining approximately accurate, would need to be slightly revised in view of stronger  degeneracy and relativistic effects. 

\begin{figure}[h]
\begin{center}
 \includegraphics[angle=0,scale=0.45]{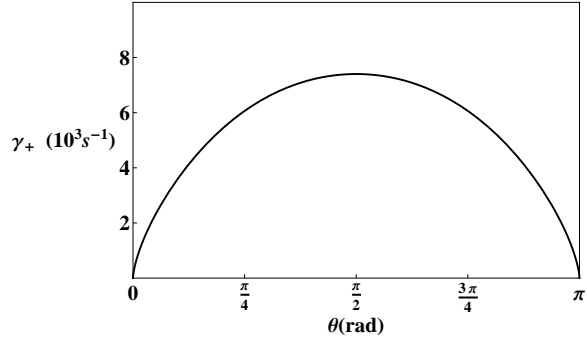}
\caption{Growth-rate of the destabilized fast magnetosonic wave, for the set of parameters described in the text.}
\label{fig2} 
\end{center}
\end{figure}

\subsection{Destabilization of the slow magnetosonic wave}  
Setting exactly the same parameters used for the fast magnetosonic wave and using Eq. (\ref{gamma}), one gets the growth-rate shown in Fig. 3 below, which is also such that $1/\gamma_- \approx 10^{-3} \,{\rm s}$. The weak beam condition (\ref{weak}) is satisfied except for 
$\theta \rightarrow \pi/2 \, {\rm rad}$, where both $\Omega_-$ and $\gamma_-$ go to zero. Contrarily to the fast magnetosonic wave, the slow magnetosonic wave becomes more unstable for parallel and anti-parallel propagation, while it stabilizes for perpendicular orientation between ${\bf k}$ and ${\bf B}_0$. 

\begin{figure}[h]
\begin{center}
 \includegraphics[angle=0,scale=0.45]{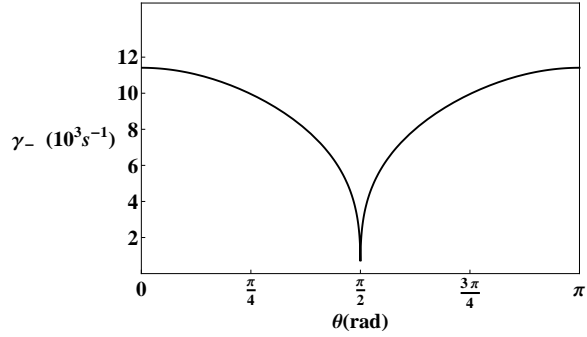}
\caption{Growth-rate of the destabilized slow magnetosonic wave, for the set of parameters described in the text.}
\label{fig3} 
\end{center}
\end{figure}

\subsection{Perpendicular wave propagation (${\bf k} \perp {\bf V}_A$)}

It is useful to collect the special cases of Eq. (\ref{gamma}) for noteworthy orientations. For instance, when $\textbf{k} \perp \textbf{B}_0$, or $\theta = \pi/2 \, {\rm rad}$, it is found 
\begin{equation}
\gamma_+ = \frac{\sqrt{3} \,\Delta^{1/3} c^{4/3} k}{2^{4/3} (V_A^2 + V_S^2)^{1/6}} \,, \quad \gamma_- = 0 \,. \label{gpl}
\end{equation}

At this point it is interesting to critically compare with the instability calculations from \cite{NMHD}, where ${\bf k} \perp {\bf B}_0$ from the beginning. There, the growth-rate was found as 
\begin{equation}
\gamma = \frac{\Delta^{1/2} c^2 k}{\sqrt{V_A^2 + V_S^2}} \,, \label{gam}
\end{equation}
see Eq. (32) therein, in the case of almost perpendicular neutrino propagation (${\bf k}\cdot{\bf u}_{\nu 0} \approx 0$), which yields the larger instabilities. While Eqs. (\ref{gpl}) for $\gamma_+$ and (\ref{gam}) for $\gamma$ are  similar, there are some decisive discrepancies, and effectively $\gamma_+ \gg \gamma$ by many orders of magnitude. This is because of the exceedingly small coupling in terms of $\Delta^{1/3} \sim G_F^{2/3}$ in Eq. (\ref{gpl}) and $\Delta^{1/2} \sim G_F$ in Eq. (\ref{gam}). What is the origin of the discrepancy? It happens that in \cite{NMHD} the neutrino-beam mode was selected with $\omega = {\bf k}\cdot{\bf u}_{\nu 0} + i \gamma$ and $\gamma \ll \Omega = (V_A^2 + V_S^2)^{1/2} k$, with wavevector almost perpendicular to neutrino beam velocity, but the resonance condition ${\bf k}\cdot{\bf u}_{\nu 0} = \Omega$ was not enforced. By definition, the resonance condition enhances the interaction between the wave and the neutrino beam, producing a larger instability. In this context the present findings are more appropriate. 

\subsection{Parallel wave propagation (${\bf k} \parallel {\bf V}_A$)}

When $\textbf{k} \parallel \textbf{B}_0$, or $\theta = 0$, we get
\begin{equation}
\gamma_+ = 0 \,, \quad \gamma_- = \frac{\sqrt{3} \,\Delta^{1/3} c^{4/3} k}{2^{4/3} V_{S}^{1/3}} \,, \label{zz}
\end{equation}
where the result supposes $V_A > V_S$. Otherwise, if $V_S > V_A$, then $\gamma_+$ is interchanged with $\gamma_-$ in Eq. (\ref{zz}). The case of parallel propagation has two fundamental modes: the pure Alfv\'en wave $\Omega = k V_A$, which is unaffected by the neutrino beam, and the sonic mode $\Omega = k V_S$, which is destabilized according to Eq. (\ref{zz}). The anti-parallel case ($\theta = \pi \,{\rm rad}$) is similar.

\section{Conclusion}

The linear dispersion relation of simplified and ideal NMHD was examined in detail, together with the validity conditions of the theory. With the additional hypothesis of very subluminal waves ($V_\pm \ll c$) and wavelengths not very small compared to the plasma skin depth, the linear dispersion relation becomes formally the same as for usual simplified and ideal MHD, provided the adiabatic sound speed is replaced by a quantity $V_S(\omega,{\bf k})$ containing the neutrino beam contribution. Therefore the standard procedure for waves with an arbitrary orientation applies. Due to the small value of the Fermi coupling constant, the neutrino term is nearly always a perturbation, to be treated as a second order effect. Nevertheless, the corresponding instability growth-rate is found to be strong enough to be a candidate for triggering cataclysmic events in supernovae. The central result of the work is the growth-rate in Eq. (\ref{gamma}), valid for arbitrary geometries and considerably enlarging the results from \cite{NMHD}, which are restricted to perpendicular wave propagation (${\bf k}\cdot{\bf B}_0 = 0$). The particular cases of destabilized fast and slow magnetosonic waves, and perpendicular and parallel propagation have been discussed. It would be interesting to relax some of the assumptions behind Eq. (\ref{eq51}), e.g. the hypotheses of very subluminal waves, as well as the introduction of non-ideality effects. In this way, even more general (and more complicated) phenomena could be addressed. 

{\bf Acknowledgments}: 
F.~H.~ acknowledges the support by Con\-se\-lho Na\-cio\-nal de De\-sen\-vol\-vi\-men\-to Cien\-t\'{\i}\-fi\-co e Tec\-no\-l\'o\-gi\-co (CNPq), and K.~A.~P.~ack\-now\-ledges the support by Coordena\c{c}\~ao de Aperfei\c{c}oamento de Pessoal de N\'{\i}vel Superior (CAPES). 


\begin{thebibliography}{99}
\bibitem{Bludman} S. Bludman, Da Hsuan Feng, Th. Gaisser and S. Pittel, Phys. Rep. {\bf 256}, 3 (1995).
\bibitem{Cooperstein} J. Cooperstein, Phys. Rep. {\bf 163}, 95 (1988).
\bibitem{Bethe1} H. A. Bethe and J. R. Wilson, Astrophys. J. {\bf 295}, 14 (1985).
\bibitem{Bethe2} H. A. Bethe, Rev. Mod. Phys. {\bf 62}, 801 (1990).
\bibitem{mag1} A. Serbeto, L. A. Rios, J. T. Mendon\c{c}a and P. K. Shukla, Phys. Plasmas {\bf 11}, 1352 (2004).
\bibitem{mag2} A. Serbeto, L. A. Rios, J. T. Mendon\c{c}a, P. K. Shukla and R. Bingham, J. Exper. Theor. Phys. (JETP) {\bf 99}, 466 (2004). 
\bibitem{mag3} R. Bingham, L. O. Silva, R. A. Cairns, V. B. Semikoz and V. N. Oraevsky, Phys. Plasmas {\bf 10}, 4903 (2003). 
\bibitem{Semikoz} V. N. Oraevsky and V. B. Semikoz, J. Exper. Theor. Phys. (JETP) {\bf 66}, 466 (2003).
\bibitem{mag5} A. J. Brizard and S. L. McGregor, New J. Phys. {\bf 4}, 97 (2002).
\bibitem{mag8} P. K. Shukla, L. Stenflo, R. Bingham, H.A. Bethe, J.M. Dawson and J.T. Mendon\c{c}a, Phys. Lett. A {\bf 230}, 353 (1997).
\bibitem{mag9} P. K. Shukla, L. Stenflo, R. Bingham, H.A. Bethe, J.M. Dawson and J.T. Mendon\c{c}a, Phys. Lett. A {\bf 224}, 239 (1997).
\bibitem{Tajima} T. Tajima and K. Shibata, {\it Plasma astrophysics} (Addison-Wesley, Reading, 1997). 
\bibitem{m1} J. T. Mendon\c{c}a, F. Haas and A. Bret, Phys. Plasmas {\bf 21}, 092117 (2014). 
\bibitem{h1} F. Haas, K. A. Pascoal and J. T. Mendon\c{c}a, Phys. Rev. E {\bf 95}, 013207 (2017).
\bibitem{h2} F. Haas, K. A. Pascoal and J. T. Mendon\c{c}a, Phys. Plasmas {\bf 24}, 052115 (2017).
\bibitem{NMHD} F. Haas, K. A. Pascoal and J. T. Mendon\c{c}a, Phys. Plasmas {\bf 23}, 012104 (2016).
\bibitem{PRD} F. Haas, K. A. Pascoal and J. T. Mendon\c{c}a,  Phys. Rev. D {\bf 96}, 023018 (2017). 
\bibitem{book} F. Haas, {\it Quantum plasmas: an hydrodynamic approach} (Springer, New York, 2011)
\bibitem{qmhd} F. Haas, Phys. Plasmas {\bf 12}, 062117 (2005).
\bibitem{Spitzer} L. Spitzer, {\it Physics of fully ionized gases}, 2nd ed. (Dover, New York, 2006).
\bibitem{Balescu} R. Balescu, {\it Transport processes in plasmas} (Elsevier, North Holland, 1988).
\bibitem{Silva} L. O. Silva, R. Bingham, J. M. Dawson, J. T. Mendon\c{c}a and P. K. Shukla, Phys. Rev. Lett. {\bf 83}, 2703 (1999). 
\bibitem{Hirata} K. Hirata, T. Kajita, M. Koshiba, M. Nakahata, Y. Oyama, N. Sato, A. Suzuki, M. Takita, Y. Totsuka, T. Kifune, T. Suda, K. Takahashi, T. Tanimori, K. Miyano, M. Yamada, E. W. Beier, L. R. Feldscher, S. B. Kim, A. K. Mann, F. M. Newcomer, R. Van, W. Zhang and B. G. Cortez, Phys. Rev. Lett. {\bf 58}, 1490 (1987). 
\end{thebibliography}
\end{document}